# Remarks on the $n = \frac{1}{2}$ disclination line in Landau-de Gennes theory of nematic liquid crystals

H. Arodź and R. Pełka

Institute of Physics, Jagellonian University,
Reymonta 4, 30-059 Cracow, Poland

**Abstract**

Using Landau-de Gennes effective theory for nematic liquid crystals we analyse structure of a rectilinear, $n = \frac{1}{2}$, smooth disclination line in the case of equal elastic constants. We find that at certain temperature there is an exact mathematical correspondence with a rectilinear vortex in superfluid $^4He$. With a help of polynomial approximation difference of free energies of smooth and singular disclination lines is estimated. It turns out that the smooth disclination line is energetically preferred only if temperature is low enough. At higher temperatures a disordered core should be expected.

PACS numbers: 61.30.Jf, 11.27.+d

# 1 Introduction

Disclination lines in nematic liquid crystals are among the most popular topological defects encountered in condensed matter physics [1], [2], [3], [4]. In spite of that, theoretical description of them still poses rather interesting problems. In the director formalism for uniaxial nematic liquid crystals there is the question of structure of a singular core of the disclination line. As it has been pointed out in papers [5], [6], [7], it is also possible to have smooth disclination lines, provided one allows for biaxiality induced by torques which are due to topologically nontrivial boundary conditions. This type of disclination lines has been investigated in the framework of Landau-de Gennes effective theory for nematic liquid crystals, mainly with the help of numerical methods. Analytic description, exact or approximate, is still missing except for special cases. Yet another theoretical description of the disclination lines can be found in [8], where Ericksen effective theory is used, or in [9] where the biaxiality is not introduced. There are also less fundamental but more difficult problems, like theoretical description of evolution of a curved disclination line, see, e.g., [10]. Actually, the present work is the first step in our attempt to solve the latter problem.

We consider the rectilinear $n = \frac{1}{2}$ disclination line within the framework of Landau-de Gennes effective theory. Review and discussion of this theory is given in, e.g., [11]. In the present paper we assume that the constant $L_2$ is equal to zero, what corresponds to equal elastic constants in the director formalism. The case of $L_2 \neq 0$ is much more complicated [12]. First, we reformulate the theoretical description of the smooth disclination line found in [7], in particular in order to prepare a convenient starting point for the analysis of the curved smooth disclination line. Next, we present certain new results. We find that there is a special case in which the transverse profile of the smooth disclination line is given by one nontrivial function, instead of two in the generic case. We provide an approximate analytic description of the disclination line by means of the polynomial approximation. Finally, comparing free energies of the smooth and singular disclination lines we notice that as we vary parameters of the model each of the two can be energetically preferred.

The contents of our paper is as follows. In Section 2 we recall formulas from Landau-de Gennes theory which we need in our considerations. Section 3 is devoted to the analytic description of the smooth rectilinear disclination line. This Section contains our new results mentioned in the preceding paragraph. In Section 4 we have collected several remarks.

## 2 Landau-de Gennes theory

The order parameter for the nematic liquid crystal in Landau-de Gennes theory has the form of a symmetric, traceless real tensor $\hat{Q} = [Q_{ij}]$, where $i, j = 1, 2, 3$. The corresponding free energy density is given by the following formulas:

$$\mathcal{F} = \frac{1}{2}L_1 \partial_i Q_{jk} \partial_i Q_{jk} + \frac{1}{2}L_2 \partial_i Q_{ik} \partial_j Q_{jk} + V(\hat{Q}), \tag{1}$$

where

$$V(\hat{Q}) = -\frac{a}{2}Tr(\hat{Q}^2) - \frac{b}{3}Tr(\hat{Q}^3) + \frac{c}{4}(Tr(\hat{Q}^2))^2. \tag{2}$$

The constants $a, b, c, L_1, L_2$ characterise the liquid crystalline material, see, e.g., [13]. In the following we assume that all the constants in (1), (2) are positive. Then, the ground states obtained by minimizing $\mathcal{F}$ are uniaxial, in accordance with what has been experimentally found for most of nematics. It is clear that in the ground state $\hat{Q}$ is constant. One can show by minimizing $V(\hat{Q})$ that all ground states $\hat{Q}_g$ can be obtained by uniform rotations or by reflections from

$$\hat{Q}_0 = \eta_0 \begin{pmatrix} 2 & 0 & 0 \\ 0 & -1 & 0 \\ 0 & 0 & -1 \end{pmatrix}, \tag{3}$$

where

$$\eta_0 = \frac{b + \sqrt{b^2 + 24ac}}{12c}. \tag{4}$$

That is,

$$\hat{Q}_g = \mathcal{O}\hat{Q}_0\mathcal{O}^T, \tag{5}$$

where $\mathcal{O} \in O(3)$ is a constant orthogonal matrix. The corresponding value of $V$ is equal to

$$V_{\min} = -\frac{3}{2}a\eta_0^2(1 + \beta), \tag{6}$$

where

$$\beta = \frac{b\eta_0}{3a}.$$

By definition, in the uniaxial case the matrix $\hat{Q}$ has two identical eigenvalues $\lambda_1 = \lambda_2 = \lambda$. Then one can define the director field $\vec{n}$ by writing $\hat{Q}$ in the form

$$\hat{Q} = 3\lambda(\frac{1}{3}I - \vec{n}\vec{n}^T), \tag{7}$$

where $I$ denotes the 3 by 3 unit matrix, the vector $\vec{n}$ has unit length, and $\vec{n}\vec{n}^T$ is a matrix (dyad). It is clear from formula (7) that as the eigenvectors corresponding to the degenerate eigenvalue $\lambda$ we can take any pair of linearly independent vectors orthogonal to $\vec{n}$. The third eigenvector is just $\vec{n}$ and the corresponding eigenvalue is equal to $-2\lambda$. For the particular $\hat{Q}_0$ given by formula (3) the director field can be taken in the form

$$\vec{n}_0 = \begin{pmatrix} 1 \\ 0 \\ 0 \end{pmatrix}, \quad \lambda = -\eta_0,$$

and
$$\vec{n}_g = \mathcal{O}\vec{n}_0 \tag{8}$$

corresponds to $\hat{Q}_g$ given by formula (5).

The true value of the Landau-de Gennes theory is of course related to the fact that it also describes states which are not the ground states. Then $\hat{Q}$ can be space and time dependent, $\hat{Q} = \hat{Q}(\vec{r}, t)$. For example, one can consider time evolution of inhomogeneous perturbations of the ground states. Another class of states contains topological defects. These states are often discussed in London's approximation [14] which consists in restricting all configurations to the set of minima of $V(\hat{Q})$, often called the vacuum manifold. Thus, in this approximation $\hat{Q}$ has the form (7) where now $\vec{n} = \vec{n}(\vec{r}, t)$ while $\lambda = -\eta_0$ remains constant. In London's approximation free energies of various states can differ only by the derivative terms in (1). Formulas (1) and (7) give the well-known Oseen-Zöcher-Frank free energy

$$\begin{aligned}\mathcal{F} = &\tfrac{1}{2}K_{11}(\mathrm{div}\vec{n})^2 + \tfrac{1}{2}K_{22}(\vec{n}\cdot\mathrm{rot}\vec{n})^2 + \tfrac{1}{2}K_{33}(\vec{n}\times\mathrm{rot}\vec{n})^2 \\ &+ \tfrac{1}{2}K_{22}\mathrm{div}[(\vec{n}\cdot\nabla)\vec{n} - \vec{n}\mathrm{div}\vec{n}] + V_{\min},\end{aligned} \tag{9}$$

where
$$K_{11} = K_{33} = 18\eta_0^2(L_1 + \frac{1}{2}L_2), \quad K_{22} = 18\eta_0^2 L_1, \tag{10}$$

and the constant $V_{\min}$ is given by formula (6). From formulas (10) we see that $L_2 = 0$ corresponds to $K_{11} = K_{22} = K_{33}$.

Unfortunately, in the case of disclination lines London's approximation is too restrictive. The point is that it allows only for disclination lines with a singular core: due to the topologically nontrivial boundary conditions $\hat{Q}$ necessarily leaves the vacuum manifold at certain points in the space. On the other hand, in the full framework of Landau-de Gennes theory also a smooth disclination line, which does not contain any singular core, is possible. Which of the two types of disclination lines is expected to occur in a concrete nematic material at a given temperature can be found out by checking the corresponding values of the free energy.

# 3 The smooth $n = \frac{1}{2}$ disclination line

## 3.1 The axially symmetric Ansatz

Topological charge pertinent to the $n = \frac{1}{2}$ disclination line is related to the fundamental group of connected components of the vacuum manifold, that is to $\pi_1(SO(3)/H)$, where $H \subset SO(3)$ is the stability group of $\hat{Q}_0$. It is a well-known fact that $\pi_1(SO(3)/H) = Z_2$, [15]. Far away from the rectilinear disclination line, that is at the spatial infinity in directions perpendicular to the line, the order parameter $\hat{Q}$ lies in the vacuum manifold (5). The matrices $\mathcal{O}$ in formula (5) taken along a large circle around the disclination line should form a continuous path in the $SO(3)$ group such that its projection on the vacuum manifold $SO(3)/H$ gives a noncontractible loop corresponding to the nontrivial element of $\pi_1(SO(3)/H)$. Let us assume that the disclination line is perpendicular to the $(x, y)$ plane. Then we may take

$$\mathcal{O}(\phi) = \begin{pmatrix} \cos\frac{\phi}{2} & -\sin\frac{\phi}{2} & 0 \\ \sin\frac{\phi}{2} & \cos\frac{\phi}{2} & 0 \\ 0 & 0 & 1 \end{pmatrix}, \qquad (11)$$

where $(\rho, \phi)$ are the polar coordinates in the $(x, y)$ plane.

The structure of the disclination line is determined from the requirement that the total free energy

$$F = \int d^3x \mathcal{F} \qquad (12)$$

has a minimum within the chosen topological class of the order parameter $\hat{Q}$. The necessary condition for that has the form

$$\frac{\delta F}{\delta Q_{ij}(\vec{x})} + \lambda_{ij} - \lambda_{ji} + \lambda\delta_{ij} = 0, \qquad (13)$$

where $\lambda_{ij}, \lambda$ are Lagrange multipliers corresponding to the conditions $Q_{ij} - Q_{ji} = 0$, $Q_{kk} = 0$, respectively. In the discussed case of equal elastic constants one may expect that the minimal free energy is obtained for an axially symmetric configuration. Therefore, we look for a smooth solution of Eq.(13) assuming the following cylindrically symmetric Ansatz

$$\hat{Q} = \frac{\eta_0}{2}\mathcal{O}(\phi)\begin{pmatrix} S(\rho) + 3R(\rho) & 0 & 0 \\ 0 & S(\rho) - 3R(\rho) & 0 \\ 0 & 0 & -2S(\rho) \end{pmatrix}\mathcal{O}^T(\phi), \qquad (14)$$

where $\eta_0$ is given by formula (4), $\mathcal{O}(\phi)$ has the form (11), and $S(\rho), R(\rho)$ are unknown functions of the polar radius $\rho$. Ansatz (14) can be written also in

the form

$$\hat{Q} = \frac{\eta_0}{2}\left[S(\rho)\begin{pmatrix} 1 & 0 & 0 \\ 0 & 1 & 0 \\ 0 & 0 & -2 \end{pmatrix} + 3R(\rho)\begin{pmatrix} \cos\phi & \sin\phi & 0 \\ \sin\phi & -\cos\phi & 0 \\ 0 & 0 & 0 \end{pmatrix}\right]. \quad (15)$$

This Ansatz is equivalent to the one considered in [7], but the form (14) is more transparent from the homotopy group viewpoint.

The appropriate boundary conditions for the functions $S(\rho), R(\rho)$ have the form

$$R(0) = 0, \quad R(\infty) = 1, \quad (16)$$

$$S(0) = w_0, \quad S(\infty) = 1, \quad (17)$$

where $w_0$ is a constant. They follow from continuity of $\hat{Q}$ at $\rho = 0$, and from the assumption that in the limit $\rho \to \infty$ $\hat{Q}$ has the form $\mathcal{O}\hat{Q}_0\mathcal{O}^T$ with $\hat{Q}_0$ given by formula (3). It is clear from the Ansatz (14) and from the conditions (16), (17) that $\hat{Q}$ is not restricted to the vacuum manifold.

The free energy density (1) is now equal to

$$\mathcal{F} = \tfrac{9}{8}a\eta_0^2\left[S'^2 + 3R'^2 + \tfrac{3}{s^2}R^2 - \tfrac{2}{3}S^2 - 2R^2 \right.$$
$$\left. -6\beta S(R^2 - \tfrac{1}{9}S^2) + \tfrac{1}{12}(1 + 3\beta)(S^2 + 3R^2)^2\right], \quad (18)$$

where ' stands for $d/ds$. For convenience we have introduced the dimensionless variable $s = \rho/\xi_0$, where $\xi_0 = \sqrt{2L_1/3a}$.

For $\hat{Q}$ of the form (15) Eq.(13) reduces to the following two equations (recall that $L_2 = 0$)

$$S'' + \frac{1}{s}S' + \frac{2}{3}S - \frac{1}{6}(1+3\beta)S(S^2 + 3R^2) - \beta(S^2 - 3R^2) = 0, \quad (19)$$

$$R'' + \frac{1}{s}R' - \frac{1}{s^2}R + \frac{2}{3}R - \frac{1}{6}(1+3\beta)R(S^2 + 3R^2) + 2\beta SR = 0. \quad (20)$$

Equations (19), (20) are quite complicated. One can notice one simple interesting solution of them besides the trivial $R = S = 0$, namely $R = 0, S = -2$. This solution describes the ground state configuration equivalent to $Q_0$. Another simple solution, namely $R = 0, S = 2/(1+3\beta)$, is not interesting because it corresponds to a local maximum of $V(\hat{Q})$.

Less trivial solutions can be studied with the help of numerical methods. In Figs.1, 2 we present two examples of numerical solutions of Eqs.(19), (20) with the boundary conditions (16), (17). The solutions are represented by the continuous lines. They were generated with the help of Maple©. The dotted lines in these figures represent approximate forms of the functions $R, S$ obtained in Subsection 3.3 below.

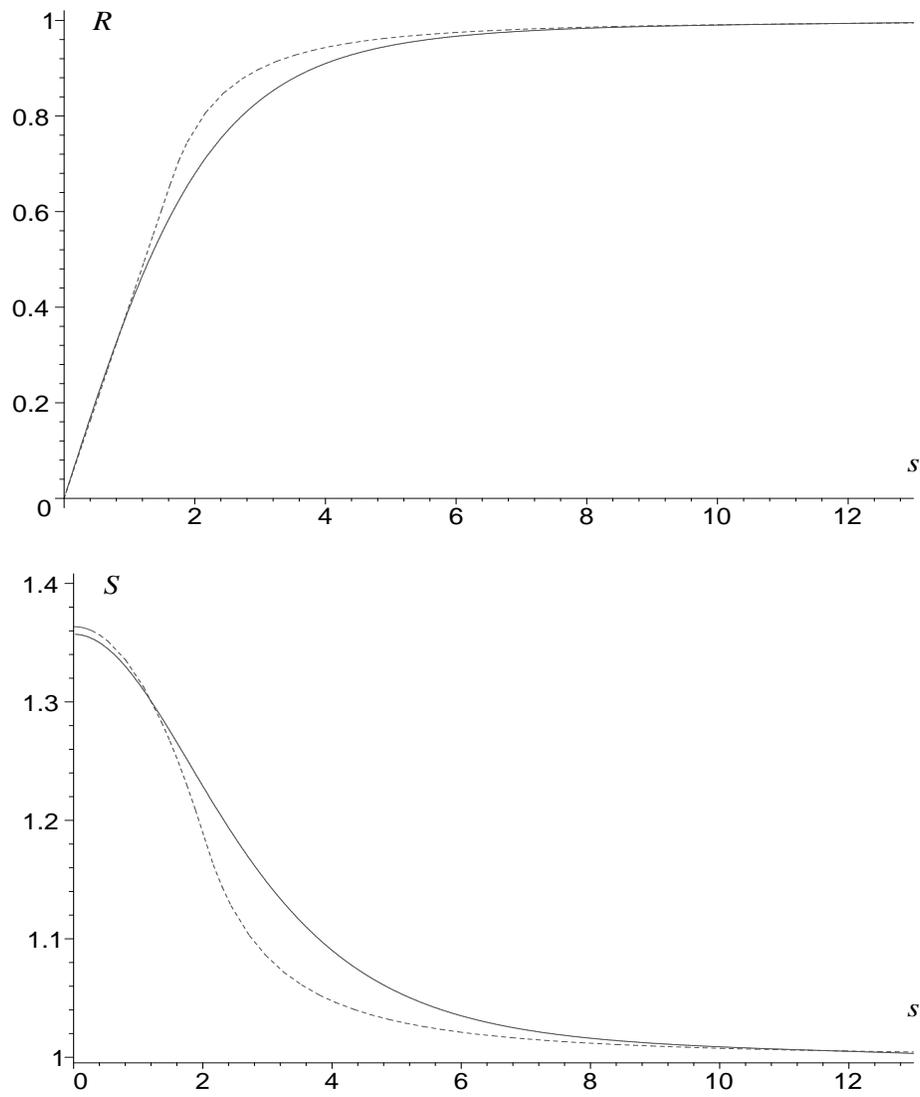

Fig.1. The functions $R$, $S$ for $\beta = 0.1$. The numerical solutions are represented by the continuous lines, while the dotted lines represent the approximate forms of $R$, $S$ constructed in Subsection 3.3. For the numerical solution $w_0 = 1.35710(4)$.

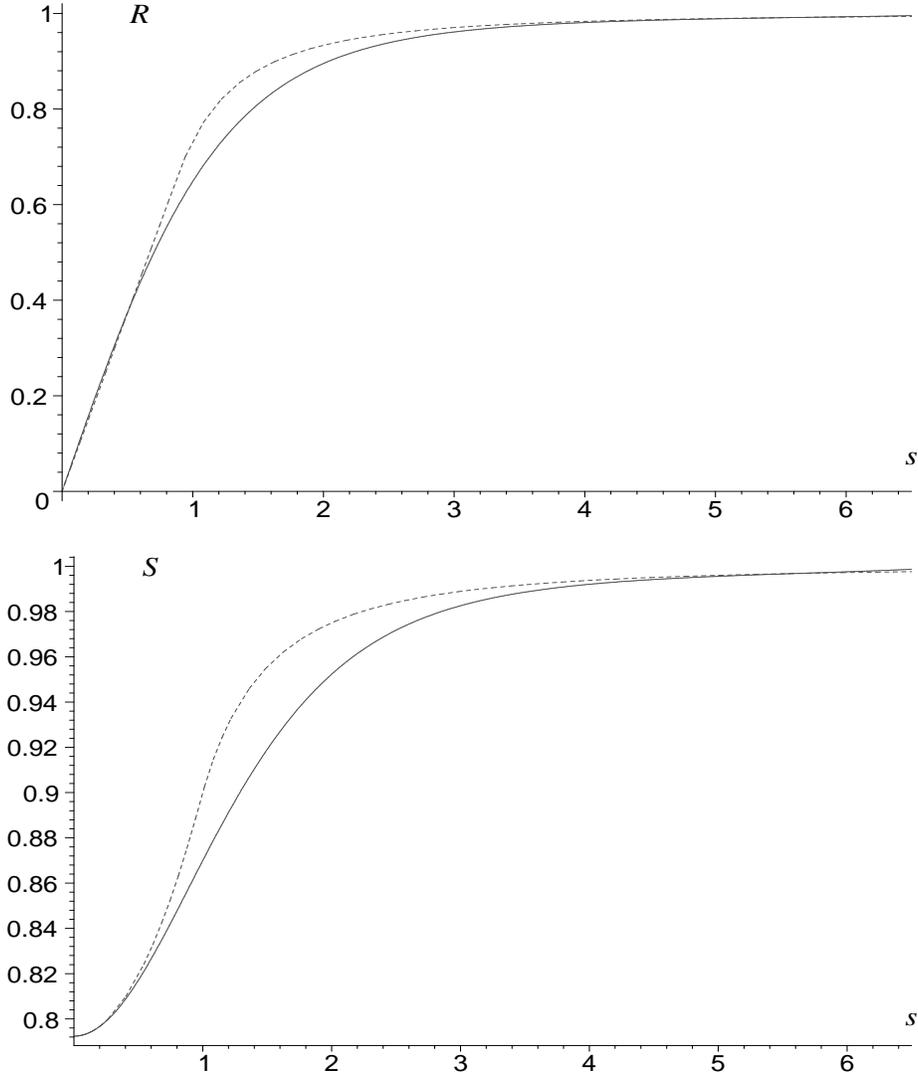

Fig.2. The functions $R$, $S$ for $\beta = 1.0$. The continuous and dotted lines have the same meaning as in Fig.1. For the numerical solution $w_0 = 0.79239(7)$.

## 3.2 The connection with Ginzburg-Pitaevskii equation

The boundary conditions (16) imply that $R(s)$ can not be constant, while for $S(s)$ this is not excluded. Actually, Figs. 1 and 2 suggest that for certain value of $\beta$ between 0.1 and 1 the function $S$ is constant. If we assume that $S =$ const, then Eq.(19) splits into two independent algebraic equations (because for $R^2$ we can substitute, e.g., 0 and 1):

$$\frac{1}{6}(1+3\beta)S - \beta = 0, \quad \frac{1}{6}(1+3\beta)S^2 + \beta S - \frac{2}{3} = 0.$$

They imply that
$$\beta = \frac{1}{3}, \quad S = 1, \tag{21}$$
and consequently
$$ac = \frac{b^2}{3}.$$
Now the remaining equation (20) has the form
$$R'' + \frac{1}{s}R' - \frac{1}{s^2}R + R(1 - R^2) = 0. \tag{22}$$

Precisely this equation appears also in the theory of superfluid $^4He$, namely Ginzburg-Pitaevskii equation for a rectilinear vortex with the unit winding number can be written exactly in the form (22), [16]. Certain similarity between the smooth disclination line and the superfluid vortex has already been noticed in [9], where purely uniaxial nematic liquid crystal with $\hat{Q}$ restricted to the form (7) is considered. Now we see that in full Landau-de Gennes theory the pertinent equations just coincide.

In the superfluid $^4He$ case the vacuum manifold can be identified with the Abelian U(1) group, hence it is quite different from $SO(3)/H$. In spite of that, the coincidence of equations suggests that dynamical properties of the supefluid vortex and of the disclination line can be quite similar.

The material constant $a$ linearly depends on temperature $T$, namely
$$a = a_0(T_* - T).$$

The nematic phase exists in certain finite temperature range,
$$T_* > T > T_m.$$

The parameter $\beta$ monotonically decreases from $\infty$ to $\beta_m$ when $T$ decreases from $T_*$ to $T_m$.

For MBBA nematic liquid crystal the material constants have the values $a_0 = 42 \times 10^3 Jm^{-3}K^{-1}, b = 64 \times 10^4 Jm^{-3}, c = 35 \times 10^4 Jm^{-3}$, and $T_* = 320K, T_m = 294K$, see, e.g., [13]. It follows from the definition of $\beta$ given below formula (6) that $\beta_m \approx 0.174$. The special value $\beta = 1/3$ corresponds to the temperature $T_{1/3} \approx 311K$. This estimate of $T_{1/3}$ for MBBA has a blemish: in reality the elastic constant $L_2 \neq 0$. Therefore, the obtained value should be taken with a grain of salt.

### 3.3 The polynomial approximation

Even for relatively simple Eq.(22) the relevant exact analytic solution is not known. Below we present an approximate solution of Eqs.(19), (20) obtained with the help of so called polynomial approximation [17]. This approximation has turned out to be very useful in several cases. It was compared with good results with purely numerical solutions in [18], where a vortex in Abelian Higgs model was considered.

We already know from Figs.1, 2 the general shape of the solutions $R$, $S$. The pictures suggest the following approximate description of these functions. For small $s$ we approximate $R$ and $S$ by the low order polynomials,

$$R_< = r_1 s, \tag{23}$$

$$S_< = w_0 + w_2 s^2, \tag{24}$$

which can be regarded as truncated series solutions of Eqs.(19), (20). Equation (19) gives the following recursive relation

$$w_2 = \frac{1}{4} w_0 \left[ \frac{1}{6}(1 + 3\beta) w_0^2 + \beta w_0 - \frac{2}{3} \right], \tag{25}$$

Here $r_1$ and $w_0$ are unknown parameters as yet. On the other hand, when $s$ is large we use the following approximate solutions of (19), (20),

$$R_> = 1 - \frac{1 + 15\beta}{12\beta(2 + 3\beta)} \frac{1}{s^2}, \quad S_> = 1 + \frac{1 - 3\beta}{4\beta(2 + 3\beta)} \frac{1}{s^2}. \tag{26}$$

The exact solutions are expected to be smooth for all $s$. Therefore, we postulate that $R_<$ smoothly matches $R_>$ at certain point $s_1$. To determine the two unknown parameters $r_1, s_1$ we may impose the two matching conditions

$$R_<(s_1) = R_>(s_1), \quad R'_<(s_1) = R'_>(s_1). \tag{27}$$

They give

$$r_1 = \frac{2}{3s_1}, \quad s_1 = \left( \frac{1 + 15\beta}{4\beta(2 + 3\beta)} \right)^{1/2}. \tag{28}$$

Similarly, for $S$ we impose the following two matching conditions at $s = s_2$

$$S_<(s_2) = S_>(s_2), \quad S'_<(s_2) = S'_>(s_2), \tag{29}$$

from which we determine $w_0$ and $s_2$. Specifically, we obtain the following relation between $w_0$ and $s_2$

$$w_0 = 1 + \frac{1 - 3\beta}{2\beta(2 + 3\beta)} \frac{1}{s_2^2}, \tag{30}$$

and the algebraic cubic equation for $w_0$

$$(1-9\beta^2)w_0^3 + 54\beta(1+\beta)w_0^2 - 4(1+21\beta+36\beta^2)w_0 + 24\beta(2+3\beta) = 0. \quad (31)$$

This last equation can easily be solved numerically for each given value of $\beta$. It is clear from formula (30) that if $0 < \beta < 1/3$ the relevant solution $w_0$ of Eq.(31) should be greater than 1, while $w_0 < 1$ for $\beta > 1/3$, and of course $w_0 = 1$ for $\beta = 1/3$. We have obtained the following approximate formulas for $s_2$ and $w_0$:
when $\beta \gg 1$

$$s_2 \approx \frac{1.171}{\sqrt{\beta}}\left(1 + \frac{0.239}{\beta} + \mathcal{O}(\beta^{-2})\right), \quad w_0 \approx 0.635 + \frac{0.190}{\beta} + \mathcal{O}(\beta^{-2}), \quad (32)$$

when $\beta \to 0$

$$s_2 \approx \frac{1}{2\sqrt{\beta}}\left(1 + \frac{15\beta}{4} + \mathcal{O}(\beta^2)\right), \quad w_0 = 2 - 12\beta + \mathcal{O}(\beta^2). \quad (33)$$

In this way we have obtained the approximate forms of $R, S$ in the full range of the $s$ variable. They are presented in Figs.1, 2 (the dotted lines) for $\beta = 0.1$ and $\beta = 1.0$. In particular, the approximate values of $w_0$ following from Eq.(31) are equal to 1.36360(9) and 0.79234(4), correspondingly. They are in quite good agreement with the values for the numerical solutions given in the figure captions.

The approximate solutions $R, S$ are continuous together with their first derivatives (the $C^1$ class) for all $s$, they converge to the exact solutions at small and large $s$, and they obey the boundary conditions (16), (17). Therefore, these approximate forms of $R, S$ give a smooth disclination line which belongs to the right topological class. Its free energy (in a finite volume) is a little bit larger than the one for the exact solutions of Eqs.(19), (20).

The Ansatz (14) and our approximate solution imply that at the center of the smooth disclination line

$$Tr(\hat{Q}^2)(s=0) = \frac{3}{2}\eta_0^2 w_0^2,$$

while in the ground state (3)

$$Tr(\hat{Q}_0^2) = 6\eta_0^2.$$

We see that degree of ordering (i.e. $Tr(\hat{Q}^2)$) inside the smooth disclination line is smaller than in the ground state. Only in the limit $\beta \to 0$, when $w_0 \to 2$, the degree of ordering approaches its ground state value.

Let us end this Subsection with two remarks about the polynomial approximation. First, there is a caveat in constructing the approximate forms of $R$ and $S$. If we include more terms in the expansions (23-26), then it turns out that always $R'_> > 0$, while the truncated series $R'_<$ has coefficients with alternating signs. Therefore, it may happen that the matching conditions (27) can not be satisfied if $R_<$ contains a wrong number of terms: at the matching point $s_1$ the highest order term in the polynomial $R'_<$ can dominate and if the sign is minus the matching is not possible at all. In the case at hand it turns out that we may take either first or fifth order polynomials, but not of the third order one. Analogously, one has to be cautious when choosing a polynomial for $S_<$.

Second, including more and more terms in the polynomials $R_<$, $S_<$ in general does not seem to be the most efficient way to improve the approximation. The point is that the infinite series expansions for $R, S$ in $s$ at small $s$, and in $s^{-1}$ at large $s$, can be slowly convergent, or even they can have a finite and too small radius of convergence. We think that it is better to take the polynomials of the lowest possible order, to construct the approximate solution, and then to compute corrections $\delta R$, $\delta S$ to it by linearising equations (19), (20) around the approximate solution. The reason is that the approximate solution constructed above already takes care of the nontrivial, nonlinear structure of the disclination line, including the boundary conditions (16), (17), and therefore we expect that $\delta R$, $\delta S$ will be small. The solutions of the resulting linear equations for $\delta R$, $\delta S$ are accessible by many techniques.

## 3.4 Free energy of the smooth disclination line

The approximate formulas for the functions $R$ and $S$ can be used to estimate the free energy per unit length of the smooth disclination line. The free energy density decreases at large $s$ rather slowly: the polynomial approximation inserted in formula (18) gives for $s \to \infty$

$$\mathcal{F} \cong V_{\min} + \frac{27a\eta_0^2}{8}\frac{1}{s^2}, \qquad (34)$$

where $V_{\min}$ is given by formula (6). This asymptotic behaviour is the same as for the singular disclination line. General shape of the free energy density can be seen from Fig.3, where we have plotted $f(s) = 8(\mathcal{F}(s) - V_{\min})/(9a\eta_0^2)$ for $\beta = 0.1$.

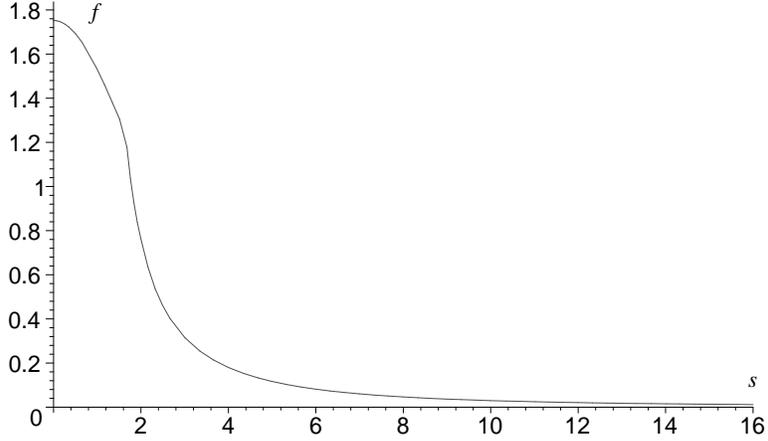

Fig.3. The normalised free energy density $f$ for $\beta = 0.1$.

Because of the slow fall off at large $s$, the total free energy per unit length of the disclination line in infinite volume is logarithmically divergent. On the other hand, the integral of the difference of the free energies densities of the smooth and singular disclination lines is convergent, but it requires an estimate of the contribution of the core of the singular disclination line to the free energy.

Let us estimate the free energy density of the singular disclination line. Outside the core the potential energy $V(\hat{Q})$ has the ground state value $V_{\min}$, and the elastic energy density given by the $L_1$ term in formula (1) has the form

$$\frac{27a\eta_0^2}{8s^2}.$$

We assume that the core is formed when the total free energy density approaches 0, because in the disordered phase $\hat{Q} = 0$ and then $\mathcal{F}$ vanishes. This means that the elastic energy at the boundary of the core is equal to $|V_{\min}|$. This condition gives the radius of the core

$$R_c = \frac{3\xi_0}{2\sqrt{1+\beta}}.$$

Furthermore, we assume that also inside the core the total free energy density vanishes. Then, the total free energy $F_c$ of the singular disclination line stored inside a concentric with it cylinder of unit height and of radius $L\xi_0$ is given by the integral

$$F_c = 2\pi\xi_0^2 \frac{9a\eta_0^2}{8} \int_{R_c/\xi_0}^{L} ds\, s(\frac{3}{s^2} - \frac{4}{3}(1+\beta)).$$

The factor $2\pi$ comes from the integration over the angle $\phi$. Note that $F_c$ includes the ground state energy $V_{\min}$.

Now we can compare the free energy $F_b$ of the smooth disclination line (inside the cylinder of radius $L\xi_0$), which is given by the formula

$$F_b = 2\pi\xi_0^2 \int_0^L dss\mathcal{F},$$

with $F_c$. For the smooth disclination line we use the polynomial approximation constructed in the previous Subsection. We have calculated numerically the difference $F_b - F_c$ for $\beta$ from 0.01 to 0.2 in steps of 0.01 taking $L=35$. The results are plotted in Fig.4.

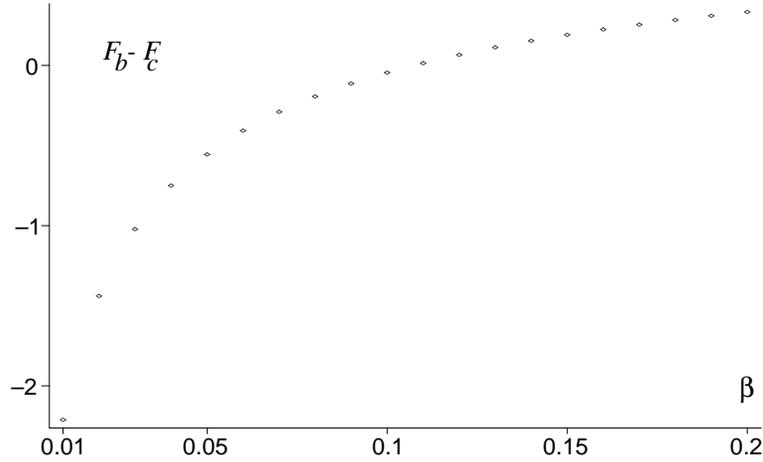

Fig.4. Difference of the free energies of the smooth disclination line ($F_b$) and of the one with the singular core ($F_c$), in units $\frac{9\pi}{4}a\eta_0^2\xi_0^2$.

We see that only for $\beta < 0.12$ the smooth disclination line has smaller total free energy. One can say that then the energetical cost of leaving the vacuum manifold is not too large. On the other hand, for larger $\beta$ it is energetically favourable to form the disordered core. Of course, our estimate of the "critical" value 0.12 of $\beta$ depends on the assumptions about the formation and energy of the core.

In Secton 3.2 we have noted that in the case of nematic MBBA $\beta > 0.174$. Therefore, our calculations suggest that in that nematic material the $n = \frac{1}{2}$ disclination line has the singular core in the whole relevant range of temperature.

# 4 Discussion

1. Let us recapitulate the main results of our work. We have found that there exists the special value $1/3$ of the parameter $\beta$ for which the smooth disclination line has particularly simple mathematical structure. We have obtained the approximate analytic formulas for the functions $R, S$. Finally, we have pointed out that for small $\beta$ the smooth disclination line is energetically preferred, while for $\beta$ large enough the singular core will appear. The critical value of $\beta$ that we have obtained is approximately equal to 0.12. This result is based on several assumptions and, therefore, should rather be regarded as an estimate only.

2. The approximate analytic description of the smooth disclination line presented in subsection 3.3 has several advantages. First, at small and large $s$ it approaches the exact solution by construction. The comparison with purely numerical solutions, see Figs.1-2, shows that it is quite good also for intermediate values of $s$. Second, it has relatively simple form and it is easy to obtain. Moreover, it gives $R, S$ in the full range $[0, \infty)$ of the independent variable $s$, as opposed to the numerical solution. Therefore, we think that in many cases the approximate analytic description can be quite a satisfactory substitute for the unknown exact analytic solution, as well as for the cumbersome purely numerical solution. Yet another argument for this comes from the fact that Landau-de Gennes model itself is also an approximation. Therefore, even the exact analytic or very precise numerical solutions of equation (13) provide only approximate description of situation in a real nematic material.

3. From physical viewpoint, perhaps the most interesting result of our paper is the suggestion that in the nematic MBBA the $n = \frac{1}{2}$ disclination lines have singular, disordered cores. However, let us remind our assumptions: $L_2 = 0$, $\hat{Q} = 0$ inside the core, the polynomial approximation for the smooth disclination line. In several papers, see, e.g., [19], structure of defects in nematics confined in spherical droplets or in a capillary has been analysed in the framework of Landau-de Gennes theory. The defects correspond to hedgehogs or to $n = 1$ disclination lines. Various possibilities for the structure of the core have been found. We think that all these results are sufficiently interesting to motivate experimental investigations of the structure of the core.

# 5  Acknowledgements

This work was supported in part by the ESF Programme "COSLAB".